# Synchronization of Fractional Chaotic Systems via Fractional-Order Adaptive Controller

S.H. Hosseinnia*, R. Ghaderi*,
A. Ranjbar N.*, J. Sadati*, S. Momani**

* *Noshirvani University of Technology, Faculty of Electrical and Computer Engineering, P.O. Box 47135-484, Babol, Iran,*
(a.ranjbar@nit.ac.ir) ,(h.hoseinnia@stu.nit.ac.ir)
** *Department of Mathematics, Mutah University, P.O. Box: 7, Al-Karak, Jordan*

**Abstract:** In this paper, an adaptive fractional controller has been designed to control chaotic systems. In fact, this controller is a fractional PID controller, which the coefficients will be tuned according to a proper adaptation mechanism. The adaptation law will be constructed from a sliding surface via gradient method. The adaptive fractional controller is implemented on a *gyro* system to signify the performance of the proposed technique.

## 1. INTRODUCTION

Fractional calculus is an old mathematical topic since 17th century. Although it has a long history, its applications to physics and engineering are just a recent focus of interest. Many systems are known to display fractional order dynamics, such as earthquake oscillation (He, 1998), Riccati (Odibat, and Momani, 2008; Cang *et. al,*. 2007), wave equation (Jafari and Momani, 2007), and chaotic equations in control engineering (Ge and Ou, 2008). There is a new topic to investigate the control and dynamics of fractional order dynamical system. The behaviour of nonlinear chaotic systems when their models become fractional have widely been investigated (Li and Chen, 2004; Ahmad and Harb Ahmad, 2003; Ahmad *et. al.*, 2007; Ahmad, 2005; Nimmo and Evans, 1999). Sensitive dependence on initial conditions is an important characteristic of chaotic systems. Therefore, chaotic systems are difficult to be synchronized or controlled. A chattering-free fuzzy sliding-mode control (FSMC) strategy for uncertain chaotic systems has been proposed in (Yau and Chen, 2006). In (Zhang *et. al.*, 2004) the authors proposed an active sliding mode control method for synchronizing two chaotic systems perturbed by parametric uncertainty. An algorithm to determine parameters of the active sliding mode controller in synchronizing different chaotic systems has been studied by (Tavazoei and Haeri, 2007). In (Yau, 2004) an adaptive sliding mode controller is presented for a class of master–slave chaotic synchronization systems with uncertainties. In (Wang and Ge, 2001) backstepping control has been proposed to synchronize the chaotic systems. Even though, synchronization has been implemented in many chaotic systems with integer derivatives, but a few works are reported on factional order chaotic system. It is because; proof of stability of the fractional order is more complex than the system with integer order. In this paper, an adaptive fractional controller has been proposed as a novel idea to control systems with fractional order dynamic. This controller is in essence a PID controller but fractional characteristics. PID coefficients $K_P$, $K_I$ and $K_D$ will be updated according to a proper gradient-based adaptation mechanism.

This paper is organized as follows:
Primarily, the proposed fractional controller will be presented in section 2, to control such similar systems. The performance of the controller will be investigated when it is used to synchronize a *gyro* dynamic. Ultimately, the work will be concluded at section 4.

## 2. FRACTIONAL ADAPTIVE CONTROLLER DESIGN

The following model represents a chaotic system with fractional order dynamic:

$$D^q x_1 = x_2$$
$$D^q x_2 = f(X,t) \qquad (1)$$

where, $0 < q \leq 1$ and $X = [x_1, x_2]^T$ is the state vector. Consider the model in (1) as a master. A secondary goal is to synchronize a usually simpler dynamic, called slave, to follow a known system, called Master. From point of view of the slave, function $f(.)$ in (1) is an unknown nonlinear function. A fractional dynamic of slave can be generally represented as:

$$D^q y_1 = y_2$$
$$D^q y_2 = f(Y,t) + \Delta f(Y,t) + d(t) + u(t) \quad (2)$$
$$, 0 < q \leq 1$$

where, $Y = [y_1, y_2]^T$ is state of the slave dynamic, $\Delta f(.)$ stands for uncertainty, $d(t)$ is disturbance and $u(t)$ is the control signal to synchronize the slave with the master. It is suggested that to synchronize via an adaptive fractional controller. The synchronization error is defined by $e_i = x_i - y_i$ where $i = 1, 2$. Schematic diagram of the closed loop system together with the proposed adaptive fractional controller is shown in Fig.(1).

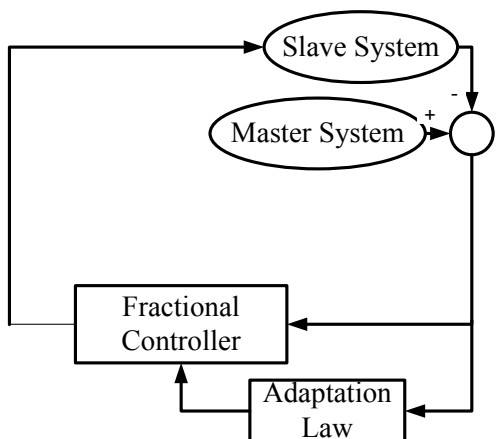

Fig.1: Schematic diagram of a synchronization mechanism

Supposed that PID controller is of the following form:
$$u(t) = K_P D^{\alpha_1} e_1(t) + K_I D^{\alpha_2} e_1(t) + K_D D^{\alpha_3} e_1(t) \quad (3)$$
It should be noted that the controller would be of the classic one if $\alpha_1 = 0, \alpha_2 = -1, \alpha_3 = 1$.

The reason behind the selection is that this kind of controller is most popular in the literature. Furthermore, the fractional controller provides the stability with more degree of freedom (Tavazoei and Haeri, 2008). To have a fractional order, parameters are chosen as $0 \leq \alpha_1 < 1$, $-1 \leq \alpha_2 < -2$ and $1 \leq \alpha_3 < 2$. Parameters of PID controller, i.e. $K_P$, $K_I$ and $K_D$ will be updated via a proper gradient-based adaptation mechanism to provide a robust synchronizing controller (Chang, 2005). The following fractional order differential equation describes a follower dummy output state $y_r$ by:

$$D^q y_r = D^q x_2 + k_2 D^q e_1 + k_1 e_1 \quad (4)$$

The sliding surface will also be defined as the error between two outputs, which is as follows:
$$S = y_2 - y_r \quad (5)$$
When the sliding mode is activated i.e. $S = 0$,

therefore we have:

$$y_2 = y_r \quad (6)$$

Since $e_2 = x_2 - y_2$ and $\dot{e}_2 = \dot{e}_1$, replacing equation (6) in (4) immediately results as:
$$D^{2q} e_1 + k_2 D^q e_1 + k_1 e_1 = 0 \quad (7)$$
Or in a state space format:
$$\begin{cases} D^q e_1 = e_2 \\ D^q e_2 = -k_1 e_1 - k_2 e_2 \end{cases} \Rightarrow D^q E = AE \quad (8)$$

where, $A = \begin{bmatrix} 0 & 1 \\ -k_1 & -k_2 \end{bmatrix}$ is the gain matrix and $E = \begin{bmatrix} e_1 \\ e_2 \end{bmatrix}$ is the error vector. If gains $k_1$ and $k_2$ are properly chosen such that the stability condition in inequality $|\arg(eig(A))| > q\pi/2$ holds, therefore the error $e_1(t)$ asymptotically tends to zero when $t \to \infty$. Let us candidate the following function as a Lyapunov function in term of the sliding surface:

$$V = \frac{1}{2} S^2 \quad (9)$$

The sliding condition will be of the form:
$$\dot{V} = S\dot{S} < 0 \quad (10)$$

When equation (10) is met, unboundedness of the sliding surface will be guaranteed when time tends to infinity. This means $S(t) \to \infty$ when $t \to \infty$. The gradient search algorithm is calculated in the direction opposite to the energy flow. Moreover, it is quite intuitive to choose $S\dot{S}$ as an error function. From equation (5) and using equation (2), we have:
$$\dot{S} = \dot{y}_2 - \dot{y}_r = D^{1-q}(D^q(y_2)) - \dot{y}_r =$$
$$D^{1-q}(f(Y,t) + \Delta f(Y,t) + d(t) + u_{PID}) - \dot{y}_r. \quad (11)$$

Pre multiplying both sides of equation (11) by $S$ yields:
$$S\dot{S} = S[D^{1-q}(f(Y,t) + \Delta f(Y,t) + d(t) + u_{PID}) - \dot{y}_r]. \quad (12)$$

Let us define the following equation:
$$U_{PID} = D^{1-q}(u_{PID}) \quad (13)$$

PID coefficients will be obtained if one uses the gradient of the adaptation law (Chang, 2005), which are as follows:

$$\dot{K}_P = -\gamma \frac{\partial S\dot{S}}{\partial K_P} = -\gamma \frac{\partial S\dot{S}}{\partial U_{PID}} \frac{\partial U_{PID}}{\partial K_P} = -\gamma S D^{1-q}(D^{\alpha_1} e_1(t)) \quad (14)$$

$$\dot{K}_I = -\gamma \frac{\partial S \dot{S}}{\partial K_I} = -\gamma \frac{\partial S \dot{S}}{\partial U_{PID}} \frac{\partial U_{PID}}{\partial K_I} = -\gamma S D^{1-q}(D^{\alpha_2} e_1(t)) \quad (15)$$

$$\dot{K}_D = -\gamma \frac{\partial S \dot{S}}{\partial K_D} = -\gamma \frac{\partial S \dot{S}}{\partial U_{PID}} \frac{\partial U_{PID}}{\partial K_D} = -\gamma S D^{1-q}(D^{\alpha_3} e_1(t)) \quad (16)$$

Substitution of $\alpha_1 = q-1$, $\alpha_2 = q-2$ and $\alpha_3 = q$ in equations (14) – (16) results the following form:

$$\dot{K}_P = -\gamma S e_1(t) \quad (17)$$

$$\dot{K}_I = -\gamma S \int_0^t e_1(\tau) d\tau \quad (18)$$

$$\dot{K}_D = -\gamma S \frac{d}{dt} e_1(t) \quad (19)$$

where, $\gamma > 0$ is the learning rate. It should be noted that $\gamma$ and $K_P$, $K_I$ and $K_D$ should be carefully selected to maintain the convergence (Chang, 2005). The proposed controller is applied on a fractional *gyro* dynamic in a synchronization task to show the performance of the technique.

## 3. SYNCHRONIZATION OF UNCERTAIN FRACTIONAL *GYRO* SYSTEM

*3.1 System description*
According to the study by Chen (Chen, 2002), dynamics of a symmetrical *gyro* with linear-plus-cubic damping of the angle $\theta$ can be expressed as (Yau, 2008):

$$\ddot{\theta} + a^2 \frac{(1-\cos\theta)^2}{\sin^3\theta} - b\sin\theta + c_1\dot{\theta} + c_2\dot{\theta}^3 \quad (20)$$
$$-\beta\sin(\omega t)\sin\theta = 0,$$

where, $\beta\sin(\omega t)$ represents a parametric excitation, $c_1\dot{\theta}$ and $c_2\dot{\theta}^3$ are linear and nonlinear damping terms, respectively, and $a^2 \frac{(1-\cos\theta)^2}{\sin^3\theta} - b\sin\theta$ is a nonlinear resilience force. Given the states $x_1 = \theta$, $x_2 = \dot{\theta}$, and

$$f(\theta,\dot{\theta}) = -a^2 \frac{(1-\cos\theta)^2}{\sin^3\theta} - c_1\dot{\theta} - c_2\dot{\theta}^3 + (b+\beta\sin(\omega t))\sin\theta,$$

this system can be transformed into the following nominal state form:

$$\begin{cases} \dot{x}_1 = x_2 \\ \dot{x}_2 = f(x_1,x_2) \end{cases} \quad (21)$$

This *gyro* system demonstrates complex dynamics. The behavior has been studied by Chen (Chen, 2002) for variety of $\beta$ in the range $32 < \beta < 36$ and constant values of $a=10, b=1$, $c_1 = 0.5$, $c_2 = 0.05$ and $\omega = 2$. Let us consider the fraction *gyro* dynamic in the following state space format:

$$\begin{cases} D^q x_1 = x_2 \\ D^q x_2 = f(x_1,x_2) \end{cases} \quad (22)$$

Fig.(2) shows the phase portrait of *gyro* chaotic system with fractional derivative in presence of $\beta = 35.5$, initial conditions of $(x_1, x_2) = (1, -1)$ and q=0.97. To show the effectiveness of the proposed controller, the procedure is implemented on fractional *gyro* dynamic.

*3.2 Implementation*

Consider system (21) as a master, which is perturbed with such an uncertainty. A slave system may be defined as the following equation:

$$\begin{cases} D^q y_1 = y_2 \\ D^q y_2 = f(y_1, y_2) + \Delta f(y_1, y_2) + d(t) + u(t) \end{cases} \quad (23)$$

Initial conditions of master and slave states are intentionally defined differently as $x_1(0) = 1$, $x_2(0) = -1$, $y_1(0) = 1.6$ and $y_2(0) = 0.8$, respectively. In order to chose an uncertainty and disturbance, $\Delta f(y_1, y_2) = -0.1\sin(y_1)$ and $d(t) = 0.2\cos(\pi t)$ are assigned, respectively. Primarily setting of PID coefficients are chosen equal to $k_p(0) = 1$, $k_I = 1$ and $K_D = 1$ and the learning rate has been selected as $\gamma = 1$. Furthermore $k_1$ and $k_2$ in (3) are selected as 0.5 and 1, respectively. Simulation results have shown in Figs. (3) to (6). In Fig. (3), synchronization of $x_1, y_1$ and $x_2, y_2$ are made perfect. The sliding surface and the control input are shown in Fig. (4) and (5) respectively, whereas Fig. (6) shows the synchronization error. It should be noted that the control signal, $u(t)$ has been activated in $t = 20s$.

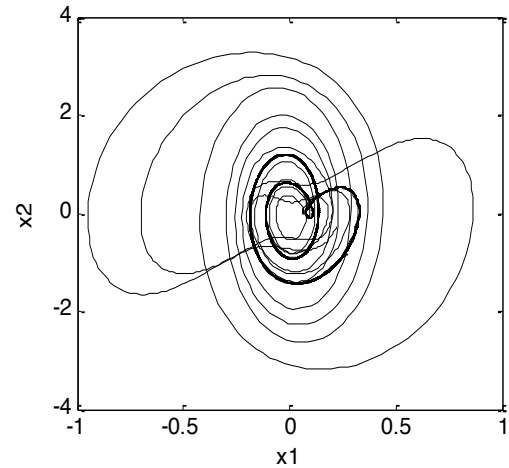

**Fig.2:** Phase portrait of fractional *gyro* chaotic system

## 4. CONCLUSION:
An adaptive fractional controller is proposed to synchronize a chaotic system. Coefficients and parameters of the controller are updated using a gradient-based adaptation mechanism. The controller has successfully been applied on the dynamic of fractional *gyro* system. The simulation results verify the significance of the proposed controller.

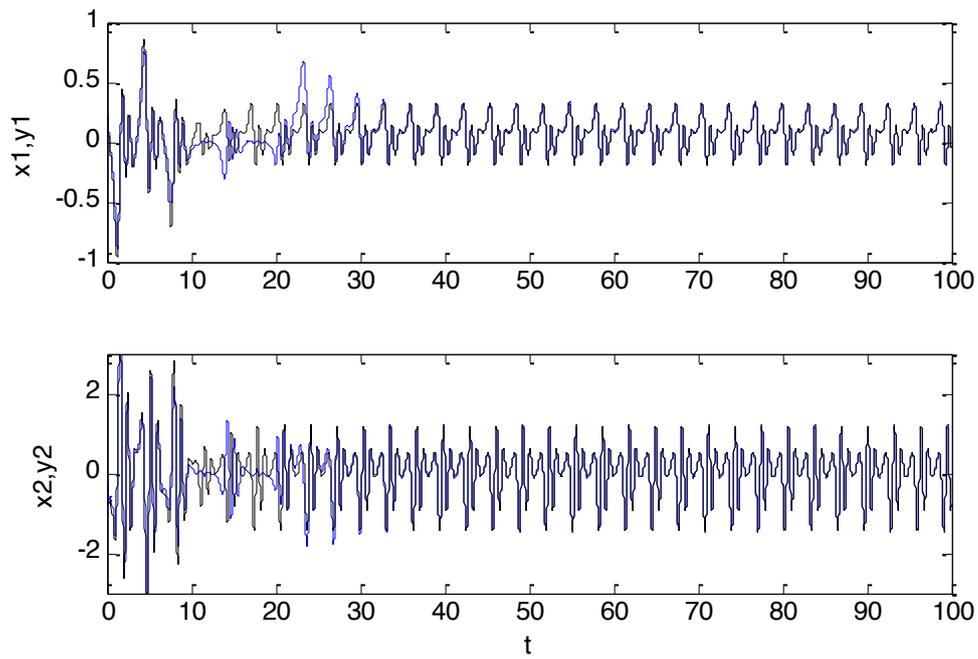

**Fig. 3:** Synchronization of Fractional *gyro*'s system

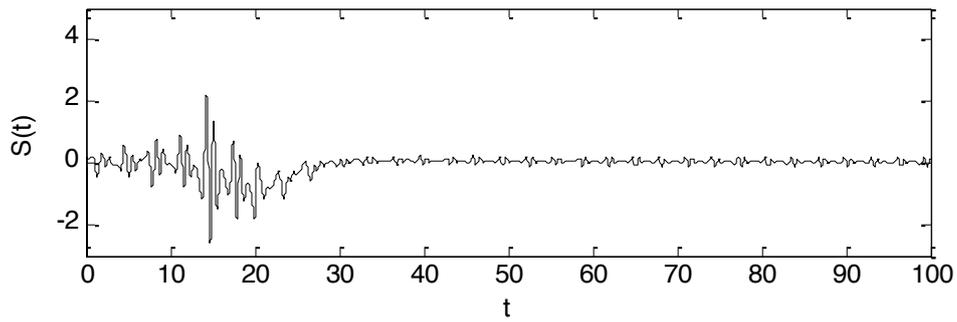

**Fig. 4:** Sliding surface

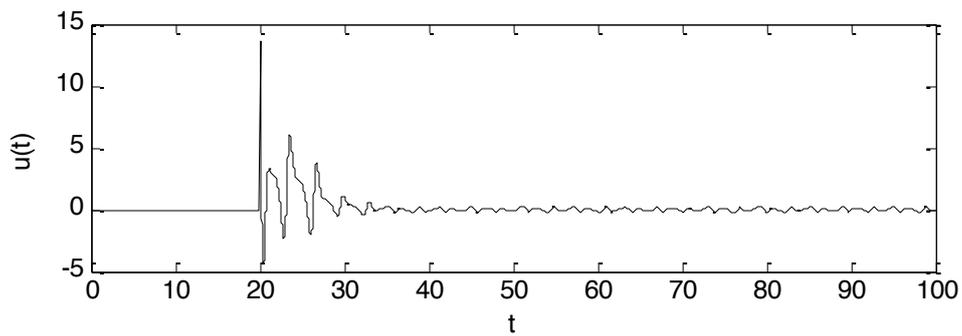

**Fig. 5:** The control signal

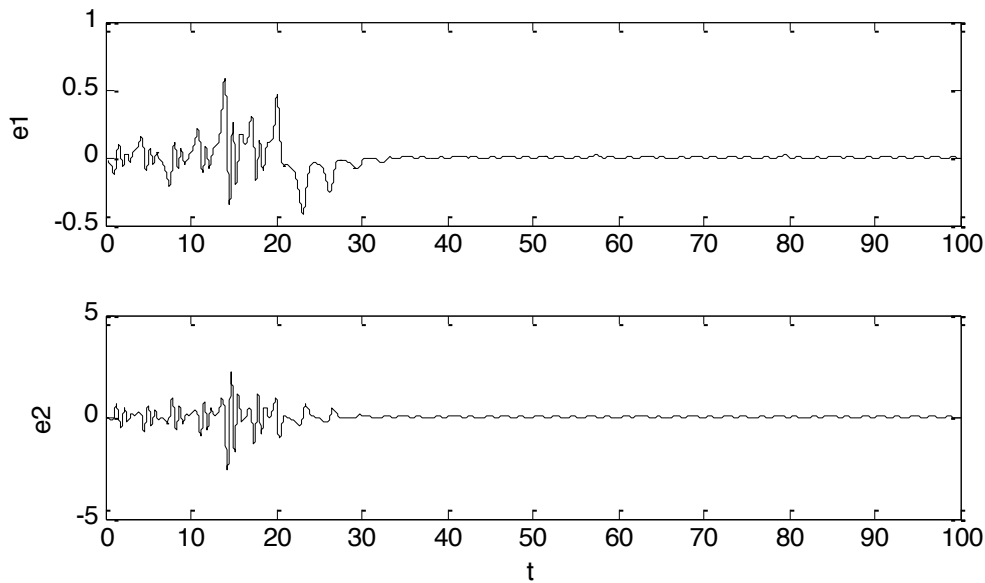

**Fig. 6:** Error of synchronization